\documentstyle[12pt]{article}
\newcommand{\be}{\begin{equation}}
\newcommand{\ee}{\end{equation}}
\newcommand{\ba}{\begin{eqnarray}}
\newcommand{\ea}{\end{eqnarray}}

\begin{document}
\begin{flushright}
Preprint SPbU-IP-94-04
\end{flushright}

\vspace{1.5cm}

\begin{center}
{\Large\bf Higher-Order Derivative SUSY \\
in Quantum Mechanics
 with Large Energy Shifts}

\vspace{0.5cm}

A. A. Andrianov\footnote{Department of Theoretical
Physics, University of Sankt-Petersburg,198904 Sankt-Petersburg, Russia.
E-mail: andrianov1@phim.niif.spb.su;\hspace{3ex}
ioffe@phim.niif.spb.su},
F. Cannata\footnote{Dipartimento di Fisica and INFN, Via Irnerio 46,
40126 Bologna, Italy. E-mail: cannata@bo.infn.it},
       and
       M. V. Ioffe$^a$
\end{center}

\vspace{1.cm}

\begin{abstract}
Within the framework of second order derivative (one dimensional)
 SUSYQM we
discuss particular realizations which incorporate large
energy shifts between the lowest states of the spectrum
of the superhamiltonian (of Schr\"odinger type). The technique
used in this construction is based on the "gluing" procedure.
We study the limit of infinite energy shift for the charges
of the Higher Derivative SUSY Algebra, and compare the results
 with  those of the standard SUSY Algebra. We conjecture that
our results can suggest a construction of a toy model where
large energy splittings between fermionic and bosonic partners
do not affect the SUSY at low energies.
\end{abstract}
\newpage

\section{Introduction}
\vspace{.5cm}
\hspace*{2ex}

First we write down for completeness the essential formalism
of one dimensional Higher Derivative SUSY Quantum Mechanics
(HSUSYQM) developed
 in \cite{AIS}
and compare it with standard SUSYQM.
The Standard SUSYQM is defined in terms of the charges $Q^{\pm}$
and the superhamiltonian $H$ with $Q^- = (Q^+)^{\dag}$
\ba
(Q^{\pm})^2 = 0,
\hspace{1cm}
[ H, Q^{\pm}] = 0, \label{basic1}\\
\hspace*{1cm} \{Q^+, Q^-\} = H \label{basic2}
\ea

The one-dimensional representation (-$\infty < x < +\infty$) is
realized by the
$2 \times 2$ supercharges
\be
Q^{+} = \left( \begin{array}{cc}
0 & 0 \\ q^- & 0 \end{array} \right)
\hspace{1cm}\mbox{and}\hspace{1cm}
Q^{-} = \left( \begin{array}{cc}
0 & q^{+} \\ 0 & 0 \end{array} \right)
\ee
where
\be
q^{\pm} = {\mp} \partial + W(x) , \quad \partial \equiv
\frac{\partial}{\partial x} \label{superp}
\ee
and
$$H = (- \partial^2 + W^2) {\bf 1} -
\sigma_3 \ W^{\prime}$$
where $\sigma_3$ is the Pauli matrix.

The spectrum of $H$ is clearly non-negative.
The Second Order extension of this algebra was recently
elaborated in \cite{AIS}. Furher investigations
concerning scattering and applications to radial problems
are represented in \cite{ACDI}.
The main difference coming from the {\em second order derivatives}
 in the charges $ q^{\pm}$
\be
q^+  = (q^-)^{\dagger}
= \partial^2 + \{ f(x), \partial\} + \phi(x), \label{gena+}
\ee
consists in the fact that the anticommutator of supercharges
generates a non Schr\"odinger type of non negative operator
$K$ :
\ba
\{Q^+, Q^-\}  =
K & =& \left(\begin{array}{cc}
k_{1} & 0 \\ 0 & k_{2} \end{array}\right) =
\left(\begin{array}{cc}
q^+ q^- & 0 \\ 0 & q^- q^+ \end{array}\right) \nonumber
\ea
such that
\ba
(Q^{\pm})^2 = 0,
\hspace{1cm}
[ K, Q^{\pm}] = 0, \label{basic11}\\
\hspace*{1cm} \{Q^+, Q^-\} = K \label{basic22}
\ea

It is instructive to try to parametrize the charges  $q^{\pm}$
 in terms of a product of {\em (first derivative)} standard
 charges\footnote
{Here we do not discuss the conditions in general which allow
this reparametrization \cite{AIS},\cite{ACDI}
but we will only consider the case in
which it is possible to solve this problem introducing the
gluing condition as discussed later.}
written in terms of
ordinary superpotentials  $W_1$ and $W_2$ (see Eq.(\ref{superp}))        ,
\be
q^+ = q_1^+ q_2^+=(-\partial + W_1(x))\, (-\partial +
W_2(x))=\partial^2+2f\partial+f^2+f^{\prime}
- W^{2} +W^{\prime} ,
\label{facta+}
\ee
\be q^- = q_2^- q_1^- = \partial^2-2f\partial+f^2-f^{\prime}
- W^{2} +W^{\prime} \label{facta-},
\ee
with the relations
\be
W_1+W_2=-2f, \quad W_1-W_2=2W, \quad \phi=f^2 - W^{2} + W^{\prime}.
\ee

In the context of a discussion of physical applications,
a relevant case is
the one in which $K$ is a (second) order polynomial
in a Schr\"odinger
like Hamiltonian operator $H$ (both operators
being 2x2 superoperators).
Such a realization is provided by the following
construction \cite{AIS},\cite{ACDI}.
Take two standard supersymmetric hamiltonians:
\be
\left(\begin{array}{cc}
h^{(1)} & 0 \\ 0 & h \end{array}\right)=\left(\begin{array}{cc}
q_1^+ q_1^- & 0 \\ 0 & q_1^- q_1^+ \end{array}\right)
\mbox{ , }
\left(\begin{array}{cc}
h & 0 \\ 0 & h^{(2)}-\lambda \end{array}\right)=\left(\begin{array}{cc}
q_2^+ q_2^- -\lambda & 0 \\ 0 & q_2^- q_2^+ -\lambda\end{array}\right),
 \label{ladder}
\ee
impose the "gluing " condition
\be
q_2^+ q_2^- -\lambda - q_1^- q_1^+ = - W_1' - W_1^2 - W_2'
+ W_2^2 -\lambda = 0, \quad \lambda>0   \label{glue}
\ee
and consider the resulting Superhamiltonian $H$  (obtained by deleting the
intermediate hamiltonian $h$):
\be
H = \left(\begin{array}{cc}
h^{(1)} & 0 \\ 0 & h^{(2)}-\lambda \end{array}\right). \label{ham}
\ee
Its hamiltonian partners $h_1,h_{2}-\lambda$  are intertwined
by the original
higher order supercharges or equivalently :
$$[ H, Q^{\pm}] = 0$$

The spectra of the two partners in $H$ are the same except for the
lowest states and the wave functions are formally connected as in
 stantard SUSYQM the supercharges being now of higher order.
In addition it turns out \cite{AIS} that the superalgebra takes the form
$$K= \{Q^+, Q^-\} = H(H+\lambda)$$

\section{Examples}
\hspace*{2ex}

We want to discuss different possibilties for the spectrum of $H$
which should exhibit a large energy shift between the lowest energy
states of the partner hamiltonians in $H$.
We discuss, in particular, cases in which the lowest eigenvalue of
$h^{(2)} - \lambda$ is $-\lambda$ so that we have to study the zero modes
 of $q^+_2$.
The general expression reads:
\be
\Psi_{- \lambda}^{(2)}(x) = \exp ( \int^x dy \ W_2(y)). \label{0mode}
\ee
The condition of normalizability can be related to the asymptotic
properties of $W_2$ for $x \rightarrow \pm \infty$:
\be
W_{2,+\infty}<0 \ ,\ W_{2,-\infty}>0. \label{topology}
\ee
More detailed information on the spectrum of $H$ also depends on $W_1$.

EXAMPLE 1.

Let us assume
$$W_1^2+W_1^{'}=\alpha^2$$
The gluing condition, Eq.(\ref{glue}), reads
$$W_2^2-W_2^{'}=\lambda+\alpha^2$$
Its general nonsingular solution is
\be
W_2(x) \equiv W_2^{(0)}(x) = -\kappa \tanh(\kappa x + \gamma),
\label{Wbase}
\ee
where $\gamma $ is arbitrary real constant and $\kappa \equiv
\sqrt{\lambda + \alpha^2}$. The asymptotic properties of
$W_2^{(0)}(x)$
$$W_{2, \pm \infty}^{(0)} = \mp \kappa$$
are in agreement with the normalizability condition,
Eq.(\ref{topology}).

Let us choose
$$ W_1=\pm\alpha $$
The spectrum of $H$ consists of a degenerate continuum spectrum
starting at $\alpha^2$. There is a non degenerate ground state
with eigenvalue $E=- \lambda$ whose wave function,
Eq.(\ref{0mode}), is:
\be
\Psi_{-\lambda}^{(2)}(x) = \cosh^{-1}(\kappa x + \gamma),
\ee
potential
$$V^{(2)}(x)=\alpha^2 - \frac{2\kappa^2}{\cosh^2(\kappa x +
\gamma)}$$
can be recognized as a familiar reflectionless potential.

A second possibility for $W_1$ is allowed i.e.
$$W_1(x)=\alpha \cdot \tanh (\alpha x)$$
The continuum spectrum is not modified but now $h^{(1)}$ acquires
a zero energy (non degenerate) bound state while the deeply bound
state of its partner as well as $V^{(2)}$ remain unmodified
\footnote{As
a remark we would like to notice that the
second possibility
of Example 1 can provide a case where $q$-deformed SUSY \cite{Spir}
techniques could
be used to generate the same partnership as the higher order derivative
charges used in the present note.

Finally we would like to make also another observation
i.e. that the same problem also can be formulated as a partnership
between
two systems associated with two hamiltonians $h^{(1)} - \alpha^2$ and
$h^{(2)} - (\lambda + \alpha^2)$, respectively. The role of
the intertwining
operators is taken now simply by the dilatation operator \cite{Spir}
(without any
further differential operator), the energies as well as the wave
function's normalization
constants turn out to be rescaled.}.

EXAMPLE 2

An alternative choice for potential  $V^{(1)}$ is given by
$$V^{(1)} = W_1^2 - W_1^{'} = \alpha^2.$$
We only study the non trivial solution (the trivial would not
be new in respect to the one given in Example 1)
\be
W_1(x)=-\alpha \cdot \tanh (\alpha x + \beta),\quad \beta = Const
\label{W1}
\ee
 for which the gluing condition Eq.(\ref{glue}) leads to
\be
W_2^2-W_2^{'}-(\lambda+\alpha^2)+
\frac{2\alpha^2}{\cosh^2(\alpha x + \beta )}=0. \label{glue1}
\ee

In this case the continuum part of the spectrum $(E \geq \alpha^2)$
of $H$ is also degenerate.
The intermediate potential
\be
V = W_1^2 + W_1^{'} = \alpha^2 - \frac{2\alpha^2}{\cosh^2(\alpha x
+ \beta)} = W_2^2 - W_2^{\prime} - \lambda   \label{glue2}
\ee
has precisely one bound state (with zero energy):
$$\Psi_{E=0}(x) = q^-_1\Psi^{(1)}_{E=0}(x).$$

Therefore operator $q_2^+q_2^-$ has the only bound state
$E = + \lambda$ (see Eq.(\ref{glue}))
but its zero-energy solution, which may be obtained from
non-normalizable function $\Psi^{(1)}_{-\lambda}(x)$,
\be
\Psi_0(x) = \exp (-\int \limits_{}^{x}W_2(y)dy) =
q^-_1 \Psi ^{(1)}_{-\lambda}(x) = (\partial + W_1)(c_1e^{\kappa x}
+ c_2e^{-\kappa x})
\label{psi}
\ee
is also non-normalizable for all constants $c_1, c_2$.

Now, by means of Darboux transformation (\ref{psi}),
we can write the general solution of the gluing equation
(\ref{glue1}):
\be
W_2(x) = -\partial \ln\Psi_0(x) = - \partial
\ln[(\partial - \alpha \tanh \alpha x + \beta)
(c_1e^{\kappa x} + c_2e^{-\kappa x})]. \label{W}
\ee

If we are interested in nonsingular solutions $W_2(x)$
it is necessary to take $c_1c_2 < 0$, i.e.
\be
W_2(x) = \alpha \tanh (\alpha x + \beta) + \frac{\lambda
\tanh (\kappa x + c)}{\alpha \tanh (\alpha x + \beta)
\tanh (\kappa x +
c) - \kappa},\quad c=const, \label{W2}
\ee
which leads to the normalizable eigenfunction of $(h^{(2)}
- \lambda)$
\be
\Psi^{(2)}_{-\lambda}(x)=\exp (+\int \limits_{}^{x}W_2(y)dy)
\label{norm}
\ee
with the eigenvalue $E=-\lambda$. Let us put the constant
$c\equiv 0$ for simplicity.
Thus the fermionic component of Hamiltonian $H$, which depends
on the real parameter $\rho$, has exactly two bound states
with wave functions
\be
\Psi_{E=0}^{(2)}(x)=(\partial + W_2) \exp(\int \limits_{}^{x}
W_1(y)dy)
\ee
and $\Psi^{(2)}_{-\lambda}(x)$ (see Eq.(\ref{norm})) where
$W_{1,2}$ were defined in Eqs.(\ref{W1}), (\ref{W2}).

For the purposes of analysis given in the next Section
let us compare the behavior of superpotentials $W_2(x)$
and $W_2^{(0)}(x)$ of the previous Examples for large
values of $\kappa$ (see Eqs.(\ref{Wbase}), (\ref{W2})).
First of all it is possible to choose the parameter
$\gamma$ in Eq.(\ref{Wbase}) such that
\be
W_2(x=0) = W_2^{(0)}(x=0). \label{zero}
\ee

At arbitrary finite $x$ and $(\mid x \mid \kappa) \gg 1 $ we
see that $$W_2(x) \sim W_2^{(0)}(x) \sim \mp \kappa$$
depending on the sign of $x$. More of that in the same
limit we have
\be
W_2(x) - W_2^{(0)}(x) \sim O(\kappa^{-1}). \label{asymp}
\ee

Before ending this section we would like to make
additional comments about the behaviour of the higher order
charges in the large $\lambda$ limit where naively one can
expect the higher order theory to become effectively the first
order (standard) SUSY because $K$ would become equal to
$H$.  What emerges from the previous examples is that for very
large values of $\lambda$ , i.e. disregarding in this limit the
very deep level , the spectrum has indeed the standard feature
of SUSYQM where one can eliminate or add or retain the lowest
bound state \footnote{ Note  however that in the scattering
(e.g.transmission coefficient) one can recognize the effects of
this state in agreement with Levinson's theorem \cite{ACDI}.}.

For a better understanding let us
consider the one dimensional scattering problem on
the line, i.e. $
x \in (- \infty, + \infty)$, for the Hamiltonians of
Eq.(\ref{ham}). The scattering wave function for
$h^{(1)}$ fulfils the asymptotic conditions
\be
\Psi^{(1)}_{k, -\infty} = e^{+ikx} + R^{(1)}(k) e^{-ikx}
\ee
and
\be
\Psi^{(1)}_{k, +\infty} = T^{(1)}(k) e^{+ikx}
\ee
where $R^{(1)}(k),\,T^{(1)}(k)$ are the reflection and transmission
coef\/f\/icients respectively.

The ladder operators $q^{\pm}$, Eq.(\ref{superp}), are
asymptotically expressed as
$$q^-_{\pm \infty} = + \partial +
W_{\pm};\hspace{1cm} q^+_{\pm \infty} = - \partial + W_{\pm}$$
with
$$W_{\pm} = \lim_{x \to \pm \infty} W(x).$$

The asymptotic scattering wave function of the partner Hamiltonian
$(h^{2}-\lambda)$ is written in the same way.
The relations between the two scattering problems for HSSQM
can be deduced by iteration from first order standard SUSY (see
for instance \cite{Cooper}):
\ba
T^{(2)}(k) = T^{(1)}(k) \frac{(k - iW_{2+})(k - iW_{1+})}
{(k - iW_{2-})(k - iW_{1-})}\nonumber\\
\hfill\nonumber\\
R^{(2)}(k) = R^{(1)}(k)\frac{(k + iW_{2-})(k + iW_{1-})}
{(k - iW_{2-})(k - iW_{1-})} \label{2scat}
\ea

One can insert the asymptotic values of the superpotentials
discussed in the previous examples and
obtain that in the large $\lambda$ limit the relations reduce
effectively to the ones of a first order theory but
for the extra
phase of $\pi$. Also the addition
or suppression of a bound state in the partner relationship
becomes transparent from the pole structure of the multiplicative
factor relating the scattering observables.

Formally we can rescale the charges $q^{\pm}$,
Eqs.(\ref{facta+}), (\ref{facta-}), by $\sqrt{\lambda}$ and
obtain the finite limiting operators
\ba
\tilde
q^- & = & \lim_{\lambda\rightarrow\infty}q^-/\sqrt{\lambda}=
\left(\lim_{\lambda\rightarrow\infty}W_2(x)/\sqrt{\lambda}\right)(
+\partial + W_1(x)) \nonumber \\ & = & -\epsilon (x)(
+\partial + W_1(x)),  \label{qtilde} \\ \tilde q^+ & =
& (\tilde q^-)^{\dagger}
= - (- \partial + W_1(x))\epsilon (x)   \nonumber
\ea
with $\epsilon(x)$ the standard step function: $\epsilon (x)=+1$
for $x>0$ and $\epsilon (x)=-1$ for $x<0$.  One can thus obtain
new super\-charges with the following algebra
\be
\{\widetilde Q^+,
\widetilde Q^-\} = H + H^{2}/\lambda  \label{algebra}
\ee
and the limiting Hamiltonian
$$\lim_{\lambda\rightarrow\infty}(H + H^2/\lambda) =
H_{\infty}$$
is unitary equivalent to the Hamiltonian of the standard
SUSYQM:
\ba
h^{(1)}_{\infty}& =& h^{(1)} = -\partial^2 + W^2_1 -
W_1^{\prime}; \nonumber \\
h^{(2)}_{\infty}& =& \epsilon (x)(- \partial^2 + W^2_1 +
W_1^{\prime})\epsilon (x). \label{infty}
\ea
Evidently the appearance of the unitary rotation $\epsilon (x)$
gives rise to the phase observed in Eqs.(\ref{2scat})
in scattering coefficients. We notice that superficially
it does not coincide with the limit
\be
\lim_{\lambda\rightarrow\infty}(h^{(2)} - \lambda) =
-\partial^2
+ W_1^2 + W_1^{\prime} - 2\sqrt{\lambda}\cdot\delta (x).
\label{delta}
\ee
Nevertheless they are equal for all $x \not= 0$ and generate
the same phase of $\pi$ in scattering characteristics.

As compared to the paper \cite{Llano} we keep track of the
existence of deep level in wave functions of $h^{(2)}_{\infty}$
and expect them to have bounce-like behavior at the origin
which however can be easily smeared out by the rotation
$\Psi^{(2)}(x) \rightarrow \epsilon (x)\Psi^{(2)}(x)$ for
$h^{(2)}_{\infty}$. Thus we impose the boundary conditions
opposite to \cite{Llano}.

\section{General Case}
\vspace{.5cm}
\hspace*{3ex}

The previous examples which we have illustrated suggest how
to formulate a general approximation scheme for arbitrary
$W_1(x)$ which by assumption is independent on $\lambda$ and
sufficiently regular at the origin.

We shall employ the Quasilinearization Method \cite{Bellman}
for the solution of gluing condition, Eq.(\ref{glue}),
$$W_2^2-W_2^{'}= W_1^2+W_1^{'}+\lambda ,$$
which can provide us with $1/\lambda$ -type expansion:
\be
W_2(x)= W_2^{(0)}(x)+W_2^{(1)}(x)+W_2^{(2)}(x)+....
\label{sum}
\ee
The procedure of solution can be started from the solution
$W_2^{(0)}(x)$, Eq.(\ref{Wbase}) of \\Example 1,
where $\alpha^2$ is now replaced by the value of
$$W_1^2(0)+W_1^{'}(0)=\rho$$
at the origin (with $\rho$ not necessarily non negative).
Thus $W_2^{(0)}(x)$ has the correct behaviour at small $x$
and the other terms in the expansion
(\ref{sum}) equal zero at the origin
by construction.

The solution for $W_2^{(1)}(x)$ is obtained
by a Quasilinearization of Eq.(\ref{glue}):
\be
2W_2^{(0)}W_2^{(1)}-W_2^{(1)'}-(W_1^2+W_1^{'}-\rho)=0,
\label{iter}
\ee
\be
W_2^{(1)}(x)= A- \exp (2 \int_0^x dy \ W_2^{(0)}(y))\left[\int_0^x(
W_1^2+W_1^{'}-\rho)\;
\exp (-2 \int_0^ydz \ W_2^{(0)}(z))dy
\right].
\label{int}
\ee
When approximating $(W_1^2+W_1^{'}-\rho)$
in the integral by a constant value $\nu$ one obtains
for $x \sqrt{\lambda}\rightarrow\pm\infty $
that $$W_2^{(1)}(x)
\rightarrow\mp\frac{\nu}{2\sqrt{\lambda+\rho}}$$
(compare this result with Eq.(\ref{asymp}) of Example 2).
Iterating this procedure to obtain $W_2^{(2)}$ one has to solve again
the same type of equation as before, Eq.(\ref{iter}):
\be
2(W_2^{(0)}+W_2^{(1)})W_2^{(2)}+(W_2^{(1)})^2-W_2^{(2)'}=0.
\ee

The resulting sum, Eq.(\ref{sum}),
shows the expected dependence on $\lambda$ of its various
terms, i.e.
$$W_2= 1/\sqrt{\lambda}[A+B/\lambda+C/\lambda^2+....].$$
In particular, the potential $V^{(2)}(x)$ can be derived in
terms of $V_1$ and $W_{1,2}$:
\be
V_2(x)=V_1(x)+2(W_1(x)+W_2(x))^{'}.
\ee
For large values of $\lambda$
$V^{(2)}(x)$ acquires an attractive  "intense"
(proportional to $\sqrt \lambda$) $\delta$-type singularity,
as in Eq.(\ref{delta}). As before the finite energy properties
of such Hamiltonian are the same for unitarily rotated regular
Hamiltonian, Eq.(\ref{infty}).

The spectrum does not in general exhibit any new features for the two
lowest bound states in respect to the ones provided by Examples 1 and 2.

\section{Conclusions}
\vspace{.5cm}
\hspace*{3ex}

The theoretical scheme we have illustrated involves an effective
non negative operator of the form $H+H^{2}/\lambda$ which belongs
to a HSUSY algebra:
$$\{\widetilde Q^+, \widetilde Q^-\} = H + H^{2}/\lambda$$

The results we have obtained may be of some relevance within
the context of effective approximation schemes for the "resolvent
operator" $1/(\lambda-H)$ (involving similar operators)
which are generally to be interpreted ($\lambda$ large) as high energy
approximations \cite{ACD}.

Extensions to higher orders (larger than two) do not seem to
present specific difficulties, leading to a Higher-order
algebra involving polynomials of $H$ \cite{AIS}, \cite{ACDI}.
They can be relevant in the approximation of relativistic
Hamiltonians in the nonrelativistic limit $(c\rightarrow
\infty)$:
$$c\sqrt{m^2c^2 + H} \sim mc^2 + H /2m - H^2 /8m^3c^2 + H^3
/16m^5c^4 ,$$
possessing the Higher-Order SUSY.

We  are thus able to provide toy models in which a finite amount
of levels is shifted down in energy in such a way that the remaining
part of the spectrum can be disregarded because it is associated
to very large energies.
However we stress that this high energy part of the spectrum
preserves the appropriate boson-fermion degeneracy important for
the ultraviolet behavior of the theory.
Note that thi[Bs mechanism for lifting the original
degeneracy is somewhat complementary to the more standard one
\cite{Nilles}
in which some levels are pushed up in energy becoming thereby
effectively unobservable. In particular we would like to stress
that our framework cannot simply be readjusted to accomodate
negative values of $\lambda$ corresponding to a positive energy
shift, this is so because the corresponding superpotentials
would become periodic and have no definite asymptotic behavior.

We have seen that the presence of deep level simulated by
$\delta$ -type potential is revealed in the discontinuity
of wave functions at the origin which can be equivalently
rotated away in the alternative description, Eq.(\ref{infty})
(our approach is different from \cite{Llano}).

{}From an algebraic point of view what we have done is to associate
the algebra of the Supercharges to some type of enveloping
algebra.  The inclusion of the quasihamiltonian $K/\lambda$
which incorporates $H^2/\lambda$ and higher order terms (if the
corresponding generalization is implemented) while preserving
SUSY requires to introduce second (higher) order charges.

The possibility to obtain a large lifting of degeneracy for the
lowest state of $H$ may be of some interest as a very simple example
where energy scales associated to fermions and bosons (as SUSY
partners) can be vastly different without necessarily breaking
a generalized HSUSY algebra.

Finally we would like to stress that the physical effects we discuss
would remain essentially unchanged in the case of $\lambda$ large but
finite thus eliminating any possible problems arising from
singularities in the limit $\lambda\rightarrow\infty$ .\\

\vspace{0.5cm}

{\large\bf  Acknowledgements}\\

We are indebted to Profs. A.Bassetto and R.Soldati
for a support of this work within the agreement
between INFN and S.Petersburg University. This work is
partially supported by GRACENAS (grant No. 1081). The
work of M.I. was partially supported by grants of
American Physical Society and Soros Foundation.

\end{document}